\documentclass[twocolumn]{aastex701}

\usepackage{amsmath} % Equations
\usepackage{enumitem}  % Lists
\usepackage{newtxtext,newtxmath} % Fonts
\usepackage{booktabs}

\begin{document}

%% Title, authors
\title{Adaptive Reconstruction of Cluster Halos (ARCH): Integrating Shear and Flexion for Substructure Detection}
\shorttitle{Adaptive Reconstruction of Cluster Halos}
\shortauthors{Shpiece \& Goldberg}

\author{Jacob Shpiece}
\affiliation{Department of Physics, Drexel University, Philadelphia, PA, USA}
\email{js4664@drexel.edu}

\author{David M. Goldberg}
\affiliation{Department of Physics, Drexel University, Philadelphia, PA, USA}
\email{dmg39@drexel.edu}

%% Abstract
\begin{abstract}
We present \textsc{ARCH} (Adaptive Reconstruction of Cluster Halos), a new gravitational lensing pipeline for cluster mass reconstruction that applies a joint shear–flexion analysis to JWST imaging. Previous approaches have explored joint shear+flexion reconstructions through forward modeling and Bayesian inference frameworks; in contrast, \textsc{ARCH} adopts a staged optimization strategy that incrementally filters and selects candidate halos rather than requiring a global likelihood model or strong priors. This design makes reconstructions computationally tractable and flexible, enabling systematic tests of multiple signal combinations within a unified framework. \textsc{ARCH} employs staged candidate generation, local optimization, filtering, forward selection, and global strength refinement, with a combined fit metric weighted by per-signal uncertainties. Applied to Abell 2744 and El Gordo, the pipeline recovers convergence maps and subcluster masses consistent with published weak+strong lensing results. In Abell 2744, the central core mass within 300 $h^{-1}$ kpc is $2.1\times10^{14}M_\odot h^{-1}$, while in El Gordo the northwestern and southeastern clumps are recovered at $2.6\times10^{14}M_\odot h^{-1}$ and $2.3\times10^{14}M_\odot h^{-1}$. Jackknife resampling indicates typical $1\sigma$ uncertainties of $10^{12}$–$10^{13}M_\odot h^{-1}$, with the all-signal and shear+$\mathcal{F}$ reconstructions providing the most stable results. These results demonstrate that flexion, when anchored by shear, enhances sensitivity to cluster substructure while maintaining stable cluster-scale mass recovery.
 \end{abstract}

%%%%%%%%%%%%%%%%% BODY OF PAPER %%%%%%%%%%%%%%%%%%

\section{Introduction}

Galaxy clusters serve as essential laboratories for advancing our understanding of cosmology and astrophysics. Their mass distributions encode information on structure growth and the properties of dark matter, while the populations of subhalos within clusters provide insight into the hierarchical assembly of galaxies and clusters. Consequently, mapping cluster mass and constraining the subhalo mass function offer dual leverage: robust tests of cold dark matter against alternative models, as well as constraints on early structure formation \citep{Vogelsberger2012,TulinYu2018,Mandelbaum2020}.

As light from background galaxies pass through galaxy clusters, gravitational interactions introduce distortions into their apparent shapes. Analysis of these distortions can recover the underlying mass distribution \citep{Clowe2006}. Shear, a first-order `stretching' effect, has long been the standard lensing observable for reconstructing large-scale mass profiles and estimating total masses \citep{Merten2011,Jauzac2016,Diego2016}. Higher-order lensing signals, namely flexion, characterize local gradients in convergence and shear: \citep{Goldberg2005,Bacon2006,Leonard2007,Okura2007}. While flexion is inherently noisier than shear, it offers enhanced sensitivity to small-scale mass fluctuations and substructures \citep{Bacon2006,Er2010,Cain2011,Cain_2016}, making it a valuable complement within dense cluster environments.

Joint reconstructions that combine multiple lensing constraints span direct inversions, forward modeling, and Bayesian frameworks \citep[e.g.][]{Bartelmann1996,bradac2005,Diego2007,Jullo2007}. In practice, most cluster-scale analyses rely primarily on shear (and, where available, strong-lensing arcs), with flexion only rarely incorporated due to measurement challenges and heterogeneous noise. The depth and angular resolution of JWST now make systematic use of flexion feasible across larger source samples, motivating methods that integrate multiple signals with appropriate weighting.

In this work we introduce \textsc{ARCH} (Adaptive Reconstruction of Cluster Halos), a reconstruction pipeline that integrates shear and flexion within a staged framework: candidate seeding, local optimization, physical/geometric filtering, forward selection with reduced $\chi^2$ improvement tests, merging of nearby candidates, and a final global strength optimization. Signal contributions are combined through a single $\chi^2$ objective that naturally weights by per-signal, per-source uncertainties. While we represent mass with a sparse set of halos (e.g. NFW) to maintain tractability, we do not impose light–traces–mass priors nor fix a single global mass model; instead, the set of retained halos is determined by the data via forward selection. We show that this approach reliably recovers known mass features in cluster environments, with a sensitivity to substructures $\gtrsim 10^{13} M_\odot h^{-1}.$

We apply \textsc{ARCH} to Abell 2744 and El Gordo, two benchmark merging clusters with extensive lensing literature \citep[e.g.]{Merten2011,Jauzac2016,Jee2014}. Our goals are to: (i) quantify how signal choices impact recovered convergence maps and substructure; (ii) compare core and subcluster masses with representative WL/WL+SL results; and (iii) assess reconstruction stability via jackknife resampling. The paper is organized as follows: Section~\ref{sec:methods} summarizes the formalism and pipeline; Section~\ref{sec:results} presents reconstructions and mass comparisons; Section~\ref{sec:discussion} discusses the roles of shear and flexion, limitations, and error characterization; and Section~\ref{sec:conclusion} concludes.

\section{Methodology}
\label{sec:methods}
\subsection{Weak Lensing Formalism}

Weak gravitational lensing describes the coherent distortions of background galaxy shapes by intervening matter along the line of sight. These distortions are traditionally decomposed into two contributions: convergence $\kappa$ and shear $\gamma$. Convergence represents the projected surface mass density in units of the critical density for lensing,  
\begin{equation}
    \begin{aligned}
        \kappa(\boldsymbol{\theta}) &= \frac{\Sigma(\boldsymbol{\theta})}{\Sigma_{\mathrm{crit}}} = \nabla^2 \psi \\
        \Sigma_{\mathrm{crit}} &= \frac{c^2}{4\pi G} \frac{D_s}{D_l D_{ls}}
    \end{aligned}
\end{equation}
where $\Sigma_{crit}$ depends on the angular diameter distances between observer, lens, and source, and where $\psi$ gives the surface gravitational potential. Convergence produces isotropic magnification of background galaxies.  

Shear induces an anisotropic stretching of galaxy images. In complex notation,  
\begin{equation}
    \gamma = \gamma_1 + i\gamma_2 
    = \frac{1}{2}\left(\partial_1^2 - \partial_2^2\right)\psi 
      + i \, \partial_1 \partial_2 \psi ,
\end{equation}
where $\partial = \partial_1 + i\partial_2$ denotes the complex gradient operator with respect to angular coordinates $(\theta_1,\theta_2)$. For a comprehensive review of convergence and shear formalism, see \citet{Bartelmann2001, Schneider2006}.  

Beyond shear, higher-order derivatives of the lensing potential generate \emph{flexion}, which describes local gradients in the lensing fields. The first flexion ($\mathcal{F}$) is the complex gradient of convergence,  
\begin{equation}
    \mathcal{F} = \partial \kappa 
    = \left( \frac{\partial \kappa}{\partial \theta_1} 
           + i \frac{\partial \kappa}{\partial \theta_2} \right),
\end{equation}
which produces a centroid shift in the direction of the lensing mass.  

The second flexion ($\mathcal{G}$) is the complex gradient of shear,  
\begin{equation}
    \mathcal{G} = \partial \gamma 
    = \left( \frac{\partial \gamma}{\partial \theta_1} 
           + i \frac{\partial \gamma}{\partial \theta_2} \right),
\end{equation}
generating a trefoil-like distortion pattern in galaxy images.

Flexion probes smaller-scale variations in the projected mass distribution than shear  \citep{Goldberg2005,Bacon2006,Leonard2007,Okura2007}. In practice, shear is sensitive to relatively smooth variations in surface mass density, while flexion responds strongly to local mass gradients. This makes flexion particularly valuable for identifying substructure within galaxy clusters, where smaller halos imprint a measurable flexion signal on nearby background sources even when their shear contribution is weak. Simulations and observational studies (e.g.\ \citep{Bacon2006, Goldberg2005, Okura2007, bird_2016, Fabritius2022}) demonstrate that flexion enhances sensitivity to sub-halo populations and improves mass reconstructions in crowded cluster environments. Its inclusion therefore allows us to probe not only the global mass distribution but also the clumpiness of matter on arcsecond scales.

Flexion is significantly noisier than shear, sensitive to galaxy morphology, PSF residuals, and pixel-level systematics \citep{Goldberg2005,Leonard2007}. {For a fixed physical flexion, $\mathcal{F}$, the distortion of a galaxy image scales with its size $a$; the dimensionless combination $\lvert a \mathcal{F} \rvert$ therefore provides a scale-invariant measure of the surface-brightness distortion. While  $\lvert a \mathcal{F} \rvert$ preserves the lensing signal under magnification, its scatter across a population of unlensed or weakly lensed galaxies is dominated by intrinsic morphological variations. This distribution offers a robust empirical estimate of the intrinsic flexion noise for sources of a given size, under the form $\frac{\sigma_{a\mathcal{F}}}{a}$\citep{Fabritius2022}.

In this work, we combine shear, first flexion, and second flexion where available. The pipeline requires at least two signals, but can incorporate all three, enabling reconstructions that exploit both the global coherence of shear and the local sensitivity of flexion.  

\subsection{Pipeline Methodology}

ARCH is a staged sequence of optimization and filtering procedures designed to balance robustness against noise with computational tractability. A single global optimization over all candidates and parameters would be intractable and prone to overfitting. Instead, we incrementally refine a deliberately overcomplete set of candidates, pruning and reoptimizing to obtain a physically plausible, statistically consistent model. 

The pipeline performs reconstructions with a combination of shear and flexion signals. At minimum two signals are required, since each lens is described by three parameters (two for position and one for strength), while each signal provides two constraints. Reconstructions can therefore be carried out with any of four combinations: $\gamma+\mathcal{F}+\mathcal{G}$, $\gamma+\mathcal{F}$, $\mathcal{F}+\mathcal{G}$, or $\gamma+\mathcal{G}$. The combined $\chi^2$ provides relative weighting among signals according to their uncertainties. Candidate lenses are represented as discrete halos, modeled either with a Singular Isothermal Sphere (SIS) or Navarro–Frenk–White (NFW) profile.  

In this work, we adopt the NFW profile as the default halo model, owing to its well-defined physical basis and analytic lensing solutions \citep{NFW1997,Bacon2006}. The NFW density profile in three dimensions is given by  
\begin{equation}
    \rho(r) = \frac{\rho_0}{(r/r_s)(1+r/r_s)^2},
\end{equation}
where $\rho_0$ is a characteristic density and $r_s$ is the scale radius. Halos are parameterized by the mass $M_{200}$ enclosed within the radius $r_{200}$, at which the mean density equals 200 times the critical density of the Universe. The concentration parameter $c_{200} \equiv r_{200}/r_s$ relates the global halo scale to the inner profile. It is more useful in a weak lensing context to compute the two dimensional projected mass density, where

\begin{equation}
    \begin{aligned}
    \Sigma(\theta) &= 2 \int_0^\infty \rho\!\left(r=\sqrt{\theta^2 + z^2} \right)\, dz \\        
    &= \frac{2 \rho_0 r_s}{x^2 - 1} f(x) \\
    f(x) &=
    \begin{cases}
        1-\frac{2}{\sqrt{1-x^2}} \mathrm{arctanh} \left( \sqrt{\frac{1-x}{1+x}}\right), & x < 1 \\
        1-\frac{2}{\sqrt{x^2-1}} \mathrm{arctan} \left( \sqrt{\frac{x-1}{1+x}}\right), & x > 1
        \end{cases}
    \end{aligned}    
\end{equation}

Where we have defined a dimensionless position coordinate $x = \theta / r_s$. From here, it is straightforward to analytically compute any of the lensing fields around some halo, given its position, mass, and concentration. 

In \textsc{ARCH} we tie the NFW concentration to halo mass and redshift via a simulation–calibrated mass–concentration relation, \(c(M_{200},z)\). We adopt the power–law form from \citet{Duffy_2008}, \(c(M,z)=A\,[M_{200}/M_0]^B(1+z)^C\), with fit parameters $A=5.71, B=-0.084, C=-0.47, M_0=2\times10^{12}M_\odot$. This form has been shown to agree with independent calibrations (e.g.\ \citealt{Maccio_2008,DuttonMaccio2014,DiemerJoyce2019}) and is widely employed in observational lensing analyses to stabilize inferences when data do not independently constrain both $M$ and $c$. In our application this reduces the per–halo parameter set from four to three (two positional coordinates plus one strength parameter, \(M_{200}\)), ensuring that with two independent lensing signals (each contributing two constraints), a single halo can be recovered without leaving the structural parameter free. The reduction in dimensionality improves robustness and tractability in crowded cluster fields and enables consistent comparisons across signal combinations, including the option to exclude a noisier channel without destabilizing the fit.

\subsubsection*{Step 1: Forecasting Lens Positions} 

 We seed one halo candidate per source position to oversample the field. This intentional oversampling seeds the pipeline without assuming the number or locations of true mass peaks. {Each candidate halo is initialized using a fast, local inversion of the measured lensing signals, placing the candidate at a source-halo separation approximated as $r\propto\gamma/\mathcal{F}$ and oriented along the flexion direction $\phi_F$. This procedure generates a spatially distributed and highly redundant set of initial candidates, ensuring sensitivity to both dominant cluster-scale halos and lower-mass substructure without imposing external priors on the mass distribution.

\subsubsection*{Step 2: Local Optimization of Candidate Lenses. }
 Each retained candidate undergoes a local three-parameter (position and strength), gradient based optimization restricted to sources within 20'' of the candidate. This radius is chosen to be large compared to the mean source separation while remaining small relative to the full field, thereby emphasizing local consistency with the surrounding lensing signal. The aim is not a global minimum but a stable local minimum that is consistent with surrounding signals, providing a reliable starting point for subsequent steps. 

\subsubsection*{Step 3: Candidate Filtering. } 
Candidates that cannot be reasonably detected by the available data are removed, on the grounds that their existence cannot be confidently predicted. These include the following criteria:
\begin{itemize}[leftmargin=*]
    \item \textbf{Physical plausibility:} Strength parameters within conservative ranges (e.g.\ $10^{10} \le M_{200} / M_\odot \le 10^{16}$) to exclude noise-dominated or unphysical solutions;
    \item \textbf{Distance-to-source constraint:} any candidate within $0.5''$ of a background source is removed, as such configurations would be expected to produce arclets or other strong-lensing features that should already have been excluded by catalog-level quality cuts.
    \item \textbf{Field-of-view constraint:} candidates are required to lie within a bounded region centered on the field, extending to $\pm 0.5$ times the full field width from the image center, to ensure they are constrained by the available data.
\end{itemize}

\subsubsection*{Step 4: Forward Selection. }
Filtered candidates are introduced sequentially into a global model. At each iteration, we test \emph{all} remaining candidates and retain only the one that yields the largest improvement in $\chi_\nu^2$; the improvement must exceed an adaptive tolerance that scales with candidate mass (reflecting expected S/N scaling). Specifically, 
a candidate of mass \(M\) is accepted only if
\[
\Delta \chi_\nu^2 > \tau(M) = \tau_0 \left( \frac{M}{M_0} \right)^p ,
\]
with \(M_0 = 10^{13}\,M_\odot\), \(p=-1\), and \(\tau_0 = 0.003\). The mass scale \(M_0\) corresponds to a characteristic group-scale halo, while the negative exponent enforces increasingly stringent acceptance criteria for higher-mass candidates, which are expected to produce stronger lensing signals. The normalization \(\tau_0\) sets the minimum improvement required for a marginally significant detection and is held fixed across all reconstructions.
Iteration stops when no candidate surpasses its tolerance, yielding a parsimonious set that meaningfully improves the fit while controlling overfitting.

\subsubsection*{Step 5: Merging of Nearby Lenses. }
Because the initial pool is redundant, multiple candidates can cluster around the same physical mass peak. Candidates separated by less than the mean source separation are merged into a single object (a standard choice approximating the effective resolution). Merging is performed via strength-weighted averaging of positions and mass. While this acts as a smoothing step, it can bias estimates if candidates with very different masses are blended; this risk is mitigated by the subsequent global reoptimization.

\subsubsection*{Step 6: Final Strength Optimization. }
Halo positions are fixed and strengths are globally reoptimized by minimizing the total $\chi^2$ of the full model. \\

Previous approaches have explored joint shear+flexion reconstructions through forward modeling \citep[e.g.][]{Cain_2016} and Bayesian inference frameworks \citep[e.g.][]{Jullo2007}. 
\textsc{ARCH} differs in adopting a staged optimization strategy that incrementally filters and selects candidate halos, rather than requiring a global likelihood model or strong priors.
This design enables reconstructions that are both computationally tractable and flexible, allowing multiple signal combinations to be tested systematically within a unified framework. 

\subsection{Lensing Catalogs}

Raw imaging data of the clusters Abell 2744 (data found at \dataset[https://doi.org/10.17909/4hd5-gn49]{https://doi.org/10.17909/4hd5-gn49}) and El Gordo (data found at \dataset[https://doi.org/10.17909/e9rr-d448]{https://doi.org/10.17909/e9rr-d448}) were obtained from the Mikulski Archive for Space Telescopes (MAST)\footnote{\url{https://mast.stsci.edu/portal/Mashup/Clients/Mast/Portal.html}} as Level~2 calibrated tiles. These tiles were processed with the JWST pipeline to produce mosaicked composite images covering the full cluster fields. From these mosaics, postage stamps centered on individual sources were extracted. Shape measurements were performed with the \textsc{Lenser}\footnote{Source code: \url{https://github.com/DrexelLenser/Lenser}.} package, which fits galaxy models simultaneously across multiple bands to provide multiband shear and flexion estimates as described below \citep{Fabritius_2020, Arena_Thesis}. Objects with poor or unstable fits were excluded based on quality metrics. The resulting filtered sample forms the weak-lensing source catalog used in subsequent analysis.

\subsubsection{Flexion Measurement and Data Cuts}
Flexion measurements were obtained with the \textsc{Lenser} pipeline, a fast, open-source Python tool with minimal dependencies. \textsc{Lenser} is a hybrid module that estimates lensing signals through an initial moment analysis, followed by local minimization of model parameters. Beyond single-band fitting, it incorporates a multi-band, multi-epoch mode in which a single galaxy model is constrained simultaneously by all available exposures across filters. For each epoch, image moments provide starting parameter estimates, with robust median values adopted across epochs. The final stage employs a global $\chi^2$ minimization, summing contributions from all bands while allowing shared structural parameters but per-band flux normalizations. This multi-band approach leverages the depth of JWST imaging to reduce band-specific noise and systematic effects, yielding more stable flexion measurements than single-band analyses. We apply \textsc{Lenser} to the JWST data release of Abell 2744 and El Gordo, implementing the multiband feature to take reliable measurements of the shear, first flexion, and second flexion. 

\subsubsection{Data Selection}

Data cuts are applied equally to each cluster, first removing any source for which \textsc{Lenser} is unable to recover a $\chi^2$ value, then applying the following standards:
\begin{itemize}[leftmargin=*]
    \item Size cuts: $0.01 < a < 2.0$ arcsec. This step removes stars, poorly resolved galaxies, and other larger artifacts.
    \item Taking $a\mathcal{F} < 0.5$, where the parameter $a\mathcal{F}$ acts as a scale invariant dimensionless scaling parameter \citep{Fabritius_2020}. This excludes outliers dominated by measurement noise. 
    \item Sèrsic radius $r_s < 5.0$ pixels, as measured by \textsc{Lenser} \citep{Sersic1963}. This condition excludes poorly resolved or blended galaxies.
\end{itemize}

After these cuts, the initial catalog of 2111 sources in Abell 2744 was reduced to 1111 usable galaxies, giving a number density of 277 per square arcminute. For El Gordo, 2714 initial detections were reduced to 1532, giving a number density of 138 per square arcminute. These filtered samples provide the basis for the reconstructions reported in Sections~\ref{abell_results} and \ref{el_gordo_results}.

\section{Results}
\label{sec:results}

We present the outcomes of applying the \textsc{ARCH} pipeline to two massive merging clusters: Abell~2744 and El~Gordo. Our analysis focuses on three key aspects: (i) the impact of source selection and data cuts, (ii) the dependence of reconstructions on signal choice, and (iii) the recovery of global mass distributions and substructures compared to literature values. We do not present results for the $\mathcal{F}+\mathcal{G}$ signal combination - while in principle this combination possesses enough information to recover the underlying mass distribution, in practice it is entirely dominated by noise.

\subsection{Abell 2744}
\label{abell_results}
The reconstructions of Abell 2744 consistently recover a dominant cluster core, with additional substructure (notably a well studied northern clump) identified in some signal combinations. Figures~\ref{fig:abell_all} and \ref{fig:abell_shearF} show representative convergence maps, while Tables~\ref{tab:abell_core} and \ref{tab:abell_north} report core and northern clump masses. All reconstructions include a mass sheet transformation such that the average value of $\kappa$ goes to zero at the image boundaries.

\begin{table}
    \centering
    \caption{Core mass of Abell 2744 within 300 $h^{-1}$ kpc for different signal combinations. Literature values are drawn from weak-lensing (WL) and joint weak+strong lensing (WL+SL) studies. The ARCH all-signal and shear+$\mathcal{F}$ runs yield core masses consistent with WL+SL estimates, while shear+$\mathcal{G}$ overshoots and $\mathcal{F}$+$\mathcal{G}$ fails entirely to locate the mass peak, underscoring the limitations of flexion-only reconstructions. Masses in units of $10^{13} M_\odot h^{-1}$.}
    \label{tab:abell_core}
    \begin{tabular}{lcc}
        \hline
        Method & Core Mass  \\
        \hline
        ARCH ($\gamma+\mathcal{F}+\mathcal{G}$) & $20.9 \pm 0.42$ \\
        ARCH ($\gamma$ + $\mathcal{F}$)   & $21.1 \pm 0.78$ \\
        ARCH ($\gamma$ + $\mathcal{G}$)   & $26.9 \pm 1.1$ \\
        \citep{Harvey2024} (WL)  & $16.0^{+0.6}_{-0.9}$  \\
        \citep{Jauzac2016} (WL+SL) & $27.7 \pm 0.1$  \\
        \citep{Merten2011} (WL+SL) & $22.4 \pm 5.5$  \\
        \citep{Medezinski2016} (WL) & $14.9 \pm 3.5$  \\
        \hline
    \end{tabular}
\end{table}

\begin{table}
    \centering
    \caption{Northern clump mass of Abell 2744 within 300 $h^{-1}$. The all-signal combination recovers a modest halo consistent with WL expectations, whereas shear+$\mathcal{F}$ fails to detect the peak entirely and shear+$\mathcal{G}$/$\mathcal{F}$+$\mathcal{G}$ substantially overestimates it. This highlights the stabilizing role of including multiple signals in tandem. Masses in units of $10^{13} M_\odot h^{-1}$.}
    \label{tab:abell_north}
    \begin{tabular}{lcc}
        \hline
        Method & Northern Clump Mass  \\
        \hline
        ARCH ($\gamma+\mathcal{F}+\mathcal{G}$) & $3.8 \pm 0.42$ \\
        ARCH ($\gamma+\mathcal{G}$)   & $16.6 \pm 1.1$ \\
        \citep{Harvey2024} (WL) & $6.5^{+0.7}_{-0.9}$ \\
        \citep{Jauzac2016} (WL+SL) & $8.6 \pm 2.2$ \\
        \hline
    \end{tabular}
\end{table}

\begin{figure}
    \centering
    \includegraphics[width=0.45\textwidth]{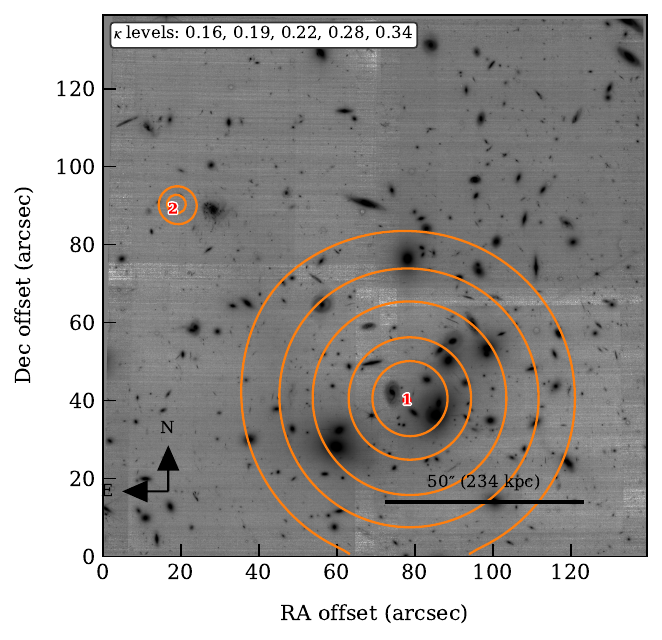}
    \caption{Convergence map of Abell 2744 reconstructed with all three signals ($\gamma, \mathcal{F}, \mathcal{G}$). The dominant central halo (labeled 1 in image) is robustly recovered, with convergence contours ($\kappa=0.16$–$0.34$) tracing the core morphology. A secondary northern clump is also detected (labeled 2 in image).}
    \label{fig:abell_all}
\end{figure}

\begin{figure}
    \centering
    \includegraphics[width=0.45\textwidth]{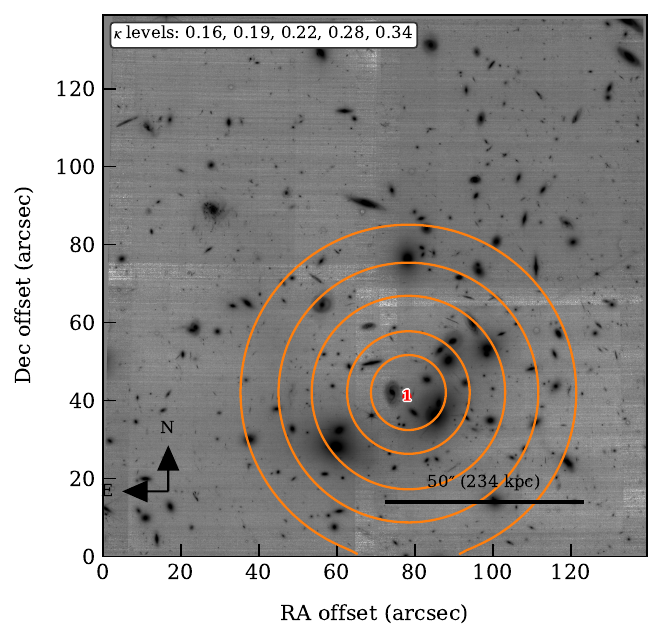}
    \caption{Convergence map of Abell 2744 reconstructed with $\gamma$ + $\mathcal{F}$. The core halo is well constrained ($\kappa=0.21$–$0.36$), consistent with the all-signal case. The northern clump is not recovered in this combination, illustrating reduced sensitivity when $\mathcal{G}$ is excluded.}
    \label{fig:abell_shearF}
\end{figure}

\begin{figure}
    \centering
    \includegraphics[width=0.45\textwidth]{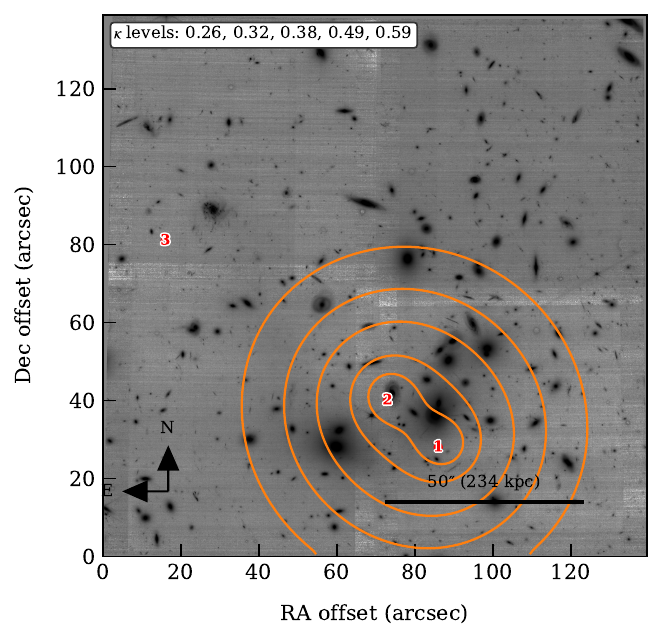}
    \caption{Convergence map of Abell 2744 reconstructed with $\gamma$ + $\mathcal{G}$. Both the core halo (peaks 1 and 2) and a northern clump (peak 3) are recovered, though the former takes a bimodal mass distribution, and the latter is overestimated relative to literature values. The inclusion of $\mathcal{G}$ improves substructure detection but introduces additional noise.}
    \label{fig:abell_shearg}
\end{figure}

\subsection{El Gordo (ACT\textnormal{-}CL~J0102\textnormal{--}4915)}
\label{el_gordo_results}
The reconstructions of El~Gordo robustly recover the well-known bimodal structure with northwestern (NW) and southeastern (SE) clumps. Figures~\ref{fig:elgordo_all} and \ref{fig:elgordo_shearflex} illustrate representative reconstructions. Mass estimates are summarized in Table \ref{tab:elgordo_subs}. To perform the mass sheet transformation for El Gordo we require $\kappa=0$ at a great distance from the image (it is insufficient to impose this condition on the image boundaries, due to the proximity of the mass peaks to said boundaries). 

In addition to the well-known NW and SE clumps, the $\gamma+\mathcal{F}+\mathcal{G}$ reconstruction (Fig.~\ref{fig:elgordo_all}) exhibits an additional peak located near the cluster center, approximately midway between the two dominant subclusters. This feature is weaker than the NW/SE halos but recurs across several signal combinations, particularly those including flexion. Its nature is ambiguous. One interpretation is that it traces a genuine intra-cluster overdensity, perhaps a group-scale halo or diffuse dark matter stripped during the ongoing merger. Numerical simulations of massive cluster collisions frequently predict transient central density enhancements as subclusters pass through one another \citep[e.g.][]{Springel2008, Vogelsberger2012}. An alternative explanation is that the apparent peak arises from the superposition of shear and flexion signals from the NW and SE clumps, which can create artificial convergence maxima in the inter-clump region. Disentangling these possibilities will require comparison with simulations and additional observational constraints, but its consistent appearance in flexion-inclusive reconstructions suggests that it may be a useful probe of small-scale dynamics in cluster mergers.

\begin{table}
    \centering
    \caption{NW and SE clump masses of El Gordo within 300 $h^{-1}$ kpc for different signal combinations. The all-signal and shear+$\mathcal{F}$ runs give mutually consistent values and reproduce the bimodal structure seen in the literature, while shear+$\mathcal{G}$ tends to inflate the SE clump and the $\mathcal{F}$+$\mathcal{G}$ run fails entirely. Masses in units of $10^{14} M_\odot h^{-1}$.}
    \label{tab:elgordo_subs}
    \begin{tabular}{lcc}
        \hline
        Method & NW & SE \\
        \hline
        ARCH ($\gamma+\mathcal{F}+\mathcal{G}$) & $2.6 \pm 0.08$ & $2.3 \pm 0.1$ \\
        ARCH ($\gamma$ + $\mathcal{F}$)   & $2.7 \pm 0.07$ & $2.3 \pm 0.08$ \\
        ARCH ($\gamma$ + $\mathcal{G}$)   & $2.8 \pm 0.04$ & $3.0 \pm 0.04$ \\
        \citep{Caminha_2023} (WL+SL) & $2.1^{+0.08}_{-0.09}$ & $2.29^{+0.09}_{-0.10}$  \\
        \citep{Diego_2023} (SL)  & 3.67 - 3.85 & 3.29 - 3.46 \\
        \hline
    \end{tabular}
\end{table}

\begin{figure}
    \centering
    \includegraphics[width=0.45\textwidth]{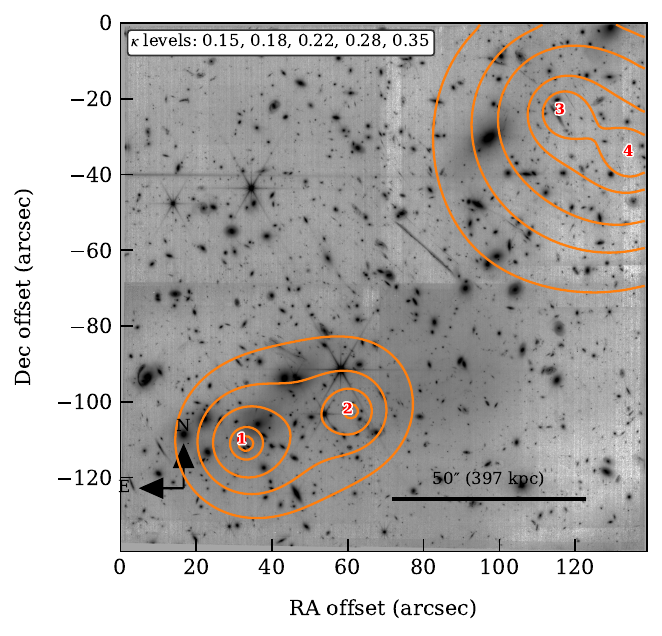}
    \caption{Convergence map of El Gordo reconstructed with all three signals. The bimodal structure is clearly recovered, with both the NW (peak 4) and SE (peak 2) clumps identified at the expected locations, albeit with spurious detections near the image boundaries, and a mass peak in the center of the cluster (peak 3). Contours follow $\kappa$ values from 0.18 to 0.37, consistent with previous weak-lensing studies.}
    \label{fig:elgordo_all}
\end{figure}

\begin{figure}
    \centering
    \includegraphics[width=0.45\textwidth]{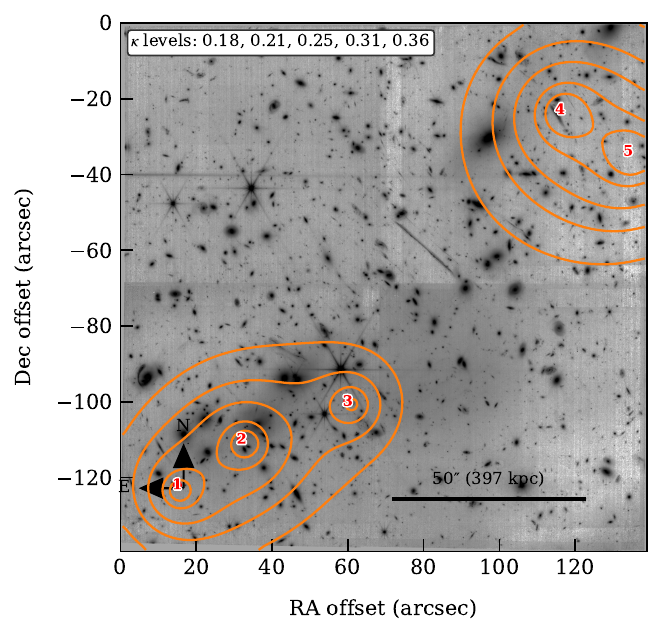}
    \caption{Convergence map of El Gordo reconstructed with $\gamma$ + $\mathcal{F}$. Both the NW and SE clumps are robustly detected, with morphologies and masses closely matching the all-signal case. This combination provides the most stable reconstruction aside from the full three-signal analysis.}
    \label{fig:elgordo_shearflex}
\end{figure}

\begin{figure}
    \centering
    \includegraphics[width=0.45\textwidth]{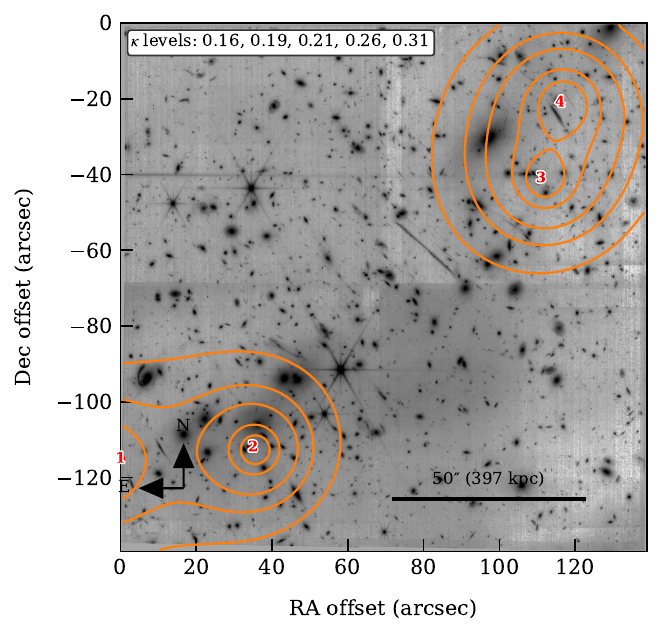}
    \caption{Convergence map of El Gordo reconstructed with $\gamma$ + $\mathcal{G}$. The bimodal structure remains visible, though the SE clump is enhanced relative to other combinations, indicating systematic amplification when $\mathcal{G}$ dominates.}
    \label{fig:elgordo_shearGflex}
\end{figure}

\subsection{Error Characterization via Jackknife Resampling}

To quantify reconstruction stability, we performed jackknife resampling in which individual sources were systematically removed from the catalogs and the full pipeline was re-executed. For each resample, the recovered masses of the principal clumps were recorded, and the statistical error was taken as the scatter across the ensemble of runs.

In Abell 2744, the central core exhibits $1\sigma$ mass uncertainties of $4.2\times10^{12}\,M_\odot h^{-1}$ (2.0\%) for the all-signal reconstruction, $7.8\times10^{12}\,M_\odot h^{-1}$ (3.7\%) for $\gamma{+}\mathcal{F}$, and $1.1\times10^{13}\,M_\odot h^{-1}$ (4.1\%) for $\gamma{+}\mathcal{G}$. The $\mathcal{F}{+}\mathcal{G}$ combination fails to yield meaningful error estimates, consistent with its instability in the main reconstructions.  

In El Gordo, uncertainties likewise depend strongly on signal choice. For the all-signal case, the NW and SE clumps are recovered with scatters of $7.8\times10^{12}\,M_\odot h^{-1}$ (3.0\% of $2.6\times10^{14}$) and $1.0\times10^{13}\,M_\odot h^{-1}$ (4.3\% of $2.3\times10^{14}$), respectively. The $\gamma{+}\mathcal{F}$ run yields comparable stability, with $7.0\times10^{12}\,M_\odot h^{-1}$ (2.6\%) for NW and $8.4\times10^{12}\,M_\odot h^{-1}$ (3.7\%) for SE. By contrast, $\gamma{+}\mathcal{G}$ shifts both masses upward but with smaller fractional scatters of $3.7\times10^{12}\,M_\odot h^{-1}$ (1.3\%) for NW and $3.6\times10^{12}\,M_\odot h^{-1}$ (1.2\%) for SE. As in Abell 2744, the $\mathcal{F}{+}\mathcal{G}$ configuration fails to converge on either clump.

Together, these results demonstrate two robust features of the ARCH pipeline. First, incorporating multiple signals reduces variance, with the all-signal and $\gamma{+}\mathcal{F}$ cases consistently achieving the most stable reconstructions. Second, the relative noisiness of $\mathcal{G}$ is reflected not only in biased mass estimates but also in larger error budgets when unaccompanied by shear. Reconstructions anchored by shear remain stable across both clusters, while flexion-only combinations are unreliable.

\section{Discussion}
\label{sec:discussion}

The ARCH pipeline systematically integrates shear, first flexion, and second flexion in a staged optimization framework. As presented above, we find that the pipeline reliably locates all major mass clumps in two well studied clusters. Here we discuss the respective contributions of shear and flexion, the robustness of the results, and the implications for future applications.  

\subsection{Cluster Results}

For Abell 2744, ARCH consistently recovers the dominant cluster core across all signal combinations. Core mass estimates from the all-signal and shear+$\mathcal{F}$ runs ($\sim 2.1 \times 10^{14}\,M_\odot h^{-1}$ within 300 $h^{-1}$ kpc) agree well with WL+SL literature values \citep[e.g.][]{Merten2011, Jauzac2016}, while the shear+$\mathcal{G}$ combination produces a modestly higher estimate and the $\mathcal{F}$+$\mathcal{G}$ combination yields an unphysically large halo. The latter emphasizes the noisiness of $\mathcal{G}$-flexion and the necessity of including shear when attempting cluster-scale reconstructions.  

The northern clump is also recovered in several configurations. The all-signal run produces a mass ($\sim 3.8 \times 10^{13}\,M_\odot h^{-1}$) consistent with literature estimates ($6.5$–$8.6 \times 10^{13}\,M_\odot h^{-1}$; \citep{Harvey2024, Jauzac2016}). By contrast, shear+$\mathcal{F}$ fails to recover this feature, while shear+$\mathcal{G}$ and $\mathcal{F}$+$\mathcal{G}$ overestimate its mass. These outcomes demonstrate that flexion contributes valuable information about substructure but must be balanced against shear to avoid overfitting.  

In El Gordo, ARCH robustly recovers the well-known bimodal configuration, with distinct NW and SE subclusters detected in the all-signal and shear+$\mathcal{F}$ runs. Aperture masses within 300 $h^{-1}$ kpc ($\sim 2.3 \times 10^{14}\,M_\odot h^{-1}$ for the SE clump and $\sim 2.6 \times 10^{14}\,M_\odot h^{-1}$ for the NW clump) are consistent with previous WL/SL studies \citep[e.g.][]{Jee2014}. Reconstructions based on shear+$\mathcal{G}$ yield somewhat elevated estimates, while $\mathcal{F}$+$\mathcal{G}$ fails to recover either subcluster robustly. These patterns mirror those in Abell 2744, reinforcing the conclusion that $\mathcal{G}$ flexion alone is too noisy for stable reconstructions, but that in combination with shear it can refine substructure estimates.  

\subsection{Flexion versus Shear Contributions}

Our reconstructions highlight complementary roles for shear and flexion in cluster lensing. Shear provides stable constraints on the global mass distribution, owing to its relatively high signal-to-noise and coherence over large angular scales \citep{Hoekstra2008,Mandelbaum2020}. This makes it well suited for recovering cluster-scale halos and establishing robust mass normalizations.  

Flexion, by contrast, probes local gradients in the potential. Its noisier character, driven by galaxy morphology and PSF residuals \citep{Goldberg2005, Leonard2007}, is offset by unique sensitivity to substructure. Prior work has shown that flexion responds strongly to subhalos and small-scale variations in surface density \citep{Bacon2006, Schneider2008, Er2010}. Our results confirm that flexion improves the localization of secondary clumps, though stable reconstructions require shear as an anchor.  

\subsection{Pipeline Robustness and Error}

The staged design of ARCH provides resilience by pruning unphysical candidates and merging redundant halos, yet residual noise sensitivity persists. In particular, $\mathcal{F}$+$\mathcal{G}$ runs are unstable, producing spurious high-mass halos. This underscores the need for weighting schemes that properly account for heterogeneous signal noise.

A further source of systematic uncertainty arises from our treatment of the mass–concentration relation. By assuming a deterministic $c_{200}(M_{200},z)$ relation \citep[e.g.][]{Ragagnin2019}, we effectively suppress the intrinsic scatter in halo concentrations at fixed mass ($\sigma_{\log c}\sim0.1$–0.2; \citealt{Ludlow2013}). This assumption reduces parameter dimensionality and stabilizes reconstructions, but may underestimate the variance of recovered halo properties, particularly in regimes where concentration scatter significantly perturbs flexion signals. Future iterations of the pipeline may wish to relax this assumption to incorporate scatter or redshift-dependent priors.

\subsection{Synthesis}

Overall, these results show that shear provides the backbone of cluster reconstructions, while flexion offers complementary small-scale sensitivity. Flexion is most effective when combined with shear, where it contributes to the detection and refinement of substructures without destabilizing the global profile. Reconstructions relying exclusively on flexion remain unstable and biased, particularly for $\mathcal{G}$-flexion. With the depth and resolution of JWST, however, flexion can now be systematically incorporated into cluster-scale analyses, offering improved constraints on subhalo masses and, by extension, the nature of dark matter.

\section{Conclusion}
\label{sec:conclusion}

We have introduced the Adaptive Reconstruction of Cluster Halos (ARCH) pipeline, a new framework designed to integrate multiple weak-lensing signals in order to probe the mass distributions of galaxy clusters. By combining shear with first- and second-order flexion, ARCH provides a flexible methodology that can operate with any two signals or all three simultaneously, balancing the complementary strengths of shear and flexion. The staged optimization strategy ensures computational tractability while maintaining sensitivity to both global cluster masses and localized substructures.

Applying ARCH to two benchmark systems---Abell 2744 and El Gordo---we have demonstrated its ability to recover robust mass reconstructions in high-density JWST fields. In Abell 2744, the pipeline reproduces a well-constrained core mass and recovers the northern clump with results broadly consistent with previous weak- and strong-lensing analyses, though the $\mathcal{F}+\mathcal{G}$ combination is shown to be unstable. In El Gordo, ARCH reliably reconstructs the well-known bimodal structure, recovering subcluster masses that agree with published values to within the spread of different signal combinations. Across both clusters, we find that the all-signal and $\gamma+\mathcal{F}$ runs provide the most stable and physically plausible results, while combinations involving $\mathcal{G}$ tend to exhibit higher variance. Future work will quantify whether the information content of $\mathcal{G}$ offsets its variance in cluster fields. 

These results highlight the value of incorporating flexion into cluster-scale weak-lensing reconstructions, particularly in the context of JWST observations, where the depth and resolution enable measurements of higher-order shape distortions at source densities not previously attainable. At the same time, they underscore the importance of careful treatment of signal choice, noise, and stability. ARCH provides a framework for carrying out such analyses systematically, extending the scope of cluster reconstructions beyond traditional shear-only methods.

Future work will extend \textsc{ARCH} to incorporate strong-lensing constraints alongside the weak-lensing signals currently used, enabling stronger constraints on recovered mass distributions. Additional directions include the use of flexion statistics to constrain the subhalo mass function and the application of multi-signal reconstructions to cosmological tests of dark matter physics. In particular, further development of multi-band flexion measurement techniques will be essential for fully exploiting the capabilities of JWST and forthcoming next-generation surveys.

\section*{Acknowledgments}
The authors would like to thank Evan J Arena and Jeimin Garibnavajwala for helpful conversations regarding the \textsc{Lenser} pipeline. 

 Some/all of the data presented in this paper were obtained from the Mikulski Archive for Space Telescopes (MAST) at the Space Telescope Science Institute. The specific observations analyzed can be accessed via \dataset[https://doi.org/10.17909/4hd5-gn49]{https://doi.org/10.17909/4hd5-gn49} and \dataset[https://doi.org/10.17909/e9rr-d448]{https://doi.org/10.17909/e9rr-d448}. STScI is operated by the Association of Universities for Research in Astronomy, Inc., under NASA contract NAS5–26555. Support to MAST for these data is provided by the NASA Office of Space Science via grant NAG5–7584 and by other grants and contracts.
%% Bibliography
\bibliographystyle{aasjournal}
\bibliography{bibtemp}

\end{document}